\documentclass[pre,twocolumn,showpacs]{revtex4}
\usepackage{amsmath}
\usepackage{amssymb}
\usepackage{graphicx}
\begin{document}

\title{NMR Experiments on a Three-Dimensional Vibrofluidized Granular Medium}

\author{Chao Huan}
\author{Xiaoyu Yang}
\author{D. Candela}
\affiliation{Physics Department, University of Massachusetts, Amherst, MA 01003}

\author{R. W. Mair}
\author{R. L. Walsworth}
\affiliation{Harvard-Smithsonian Center for Astrophysics, Cambridge, MA 02138}

\begin{abstract}
	A three-dimensional granular system fluidized by vertical container vibrations was studied using pulsed field gradient (PFG) NMR coupled with one-dimensional magnetic resonance imaging (MRI).
	The system consisted of mustard seeds vibrated vertically at 50~Hz, and the number of layers $N_\ell \leq 4$ was sufficiently low to achieve a nearly time-independent granular fluid.
	Using NMR, the vertical profiles of density and granular temperature were directly measured, along with the distributions of vertical and horizontal grain velocities.
	The velocity distributions showed modest deviations from Maxwell-Boltzmann statistics, except for the vertical velocity distribution near the sample bottom which was highly skewed and non-Gaussian.
	Data taken for three values of $N_\ell$ and two dimensionless accelerations $\Gamma=15,18$ were fit to a hydrodynamic theory, which successfully models the density and temperature profiles including a temperature inversion near the free upper surface.
\end{abstract}

\date{12 May 2003}
\maketitle

\section{\label{secintro}Introduction}

	One of the basic granular flow phenomena is the creation of a fluidized state by vibration of the container that holds the granular medium.
	If the container is vibrated vertically with a sinusoidal waveform $z(t) = z_0 \cos(\omega t)$, the dimensionless acceleration amplitude relative to gravity is $\Gamma = \omega^2 z_0/g$ and $\Gamma \agt 1$ is required to set the granular system into motion \cite{poschel00}.
	A second key dimensionless parameter is $X = N_\ell(1-\epsilon)$, where $\epsilon$ is the velocity restitution coefficient for grain-grain collisions~\cite{noteepsilon} and $N_\ell$ is the number of grain layers (the height of the granular bed at rest divided by the grain diameter).
	Experiments and simulations in one and two dimensions suggest that $X < X_f$ with $X_f \sim 1$ is required to reach a \emph{uniformly fluidized} state, in which the stochastic motions of the grains exceed motions that are harmonically or sub-harmonically related to the container vibration \cite{luding94a,luding94,bernu94}.
	For larger $X > X_f$ inelastic collisions rapidly quench stochastic grain motion and the granular bed moves as a coherent mass over much of the vibration cycle, leading to phenomena such as subharmonic surface patterns and localized ``oscillon'' structures \cite{melo95,clement96,umbanhowar96}.

	In this paper we report an experimental study of the uniformly fluidized state in three dimensions, using NMR techniques to probe the grain density profile and velocity distribution.
	Many authors have applied the methods of statistical mechanics to the fluidized granular state \cite{jenkins83,haff83,lun84,grossman97,vannoije98,sela98,garzo99,soto01}.
	Typically, one defines a ``granular temperature'' proportional to the random kinetic energy per degree of freedom of the grains, and hopes to apply hydrodynamic concepts such as thermal conductivity and viscosity to the granular fluid.
	Using NMR we are able to directly measure the granular temperature profile in a dense, three-dimensional sample for comparison with hydrodynamic theory.

	Such comparisons are significant as the applicability of hydrodynamics to granular systems is questionable.
	In a vibrofluidized state with energy injection from below, as studied here, the granular temperature varies sharply with height over distances comparable to the mean free path.
	Hence the separation between microscopic and macroscopic (hydrodynamic) length scales is marginal, and hydrodynamics may or may not provide an accurate description of the system \cite{tan98,dufty99,goldhirsch01}.

	Previous experiments have created (quasi) \emph{two-dimensional} systems by confining grains between vertical plates or on a horizontal or tilted surface, enabling data collection by high-speed video \cite{clement91,warr95,losert99,olafsen99,wildman99,rouyer00,blair01,rericha01}.
	\emph{Three-dimensional} granular systems have been studied experimentally by external probes such as capacitance \cite{knight95}, mutual inductance \cite{falcon99a} and diffusing-wave spectroscopy \cite{menon97}, as well as by non-invasive internal probes including NMR \cite{nakagawa93,knight96,caprihan97,yang00,mueth00,caprihan00} and positron-emission particle tracking (PEPT) \cite{wildman01b}.
	The PEPT method has been used to measure the granular temperature profile of a vibrofluidized bed at volume filling fractions up to 0.15 \cite{wildman01b}.
	In the present study using NMR, we have measured granular temperature profiles at much higher densities, up to 0.45 volume filling fraction.
	Density is a key parameter for granular flows, as studies suggest the existence of long-lived fluctuations and glassy, out-of-equilibrium behavior as the density increases towards the random-close-packed volume filling fraction $\approx 0.6$, signaling a breakdown of conventional hydrodynamics \cite{denniston99,howell99,silbert02,longhi02}.
	Even in the absence of glassiness there are precollision velocity correlations which increase with density \cite{soto01,pagonabarraga01}; these correlations are included in some hydrodynamic theories \cite{vannoije98} but ignored in others (``molecular chaos'' assumption) \cite{garzo99}.

	Other workers have used NMR to study the dense granular flows that occur in a rotating drum~\cite{nakagawa93,caprihan00}, for quasi-static tapping~\cite{knight96}, in a period-doubled arching flow~\cite{caprihan97}, and in a Couette shear apparatus~\cite{mueth00}.  In contrast to these earlier studies we probe a system sufficiently simple to be described by first-principles granular hydrodynamics.  In addition, we use rapid gradient pulse sequences to probe deep within the ballistic regime, which has not been possible for these very dense systems.

	In the experiments described here, small samples consisting of 45 - 80 mustard seeds of mean diameter 1.84 mm were confined to a cylindrical sample tube with 9 mm inside diameter, and vibrated vertically at $\omega/2\pi = 50$~Hz, $\Gamma =15$--18.
	The sample container was sufficiently tall to prevent grain collisions with the container top.
	The total number of grains is small in our experiments (compared, for example, with the number of particles in recent molecular-dynamics simulations \cite{bougie02}).
	However these physical experiments include effects present in practical applications but not usually addressed by simulations: irregular and variable grain size, grain rotation, realistic inelastic collisions, and the effect of interstitial gas.

	The granular systems studied here were small both horizontally (5 grain diameters) and vertically ($N_\ell \le 4$), and thus it may be surprising that they can be successfully described by a hydrodynamic theory.
	The small sample diameter was simply a consequence of the inner diameter of the NMR apparatus that was available, and could be increased in future studies.
	Conversely, the sample depth is inherently limited to a small number of layers by the fluidization condition $N_\ell(1-\epsilon) \alt 1$.
	Therefore \emph{physical} vibrofluidized systems (which typically have $\epsilon \leq 0.9$) are limited to $N_\ell \alt 10$ and the applicability of granular hydrodynamics is a nontrivial question to be tested by experiment.

	A short report of some of this work was published in Ref.~\onlinecite{yang02}, using a single set of $N_\ell, \Gamma$ values.
	For the present report, both $N_\ell$ and $\Gamma$ were varied. In addition improved methods were developed for calibrating the NMR gradient coils, resulting in more accurate granular temperature profiles.
	In particular, better agreement is found with hydrodynamic predictions for the ``temperature inversion'' that occurs at the free upper surface.

\section{\label{expt}Experimental methods}

\subsection{Experimental setup for NMR of vibrated media}

	The experiments were carried out in a vertical-bore superconducting magnet with a lab-built 40 MHz NMR probe equipped with gradient coils for pulsed field gradient (PFG) and magnetic resonance imaging (MRI) studies.
	The low static field value (0.94~T) was chosen to minimize magnetic field gradients in the sample due to the susceptibility difference between grains and voids.
	This permitted the use of a simple unipolar PFG sequence (Fig.~\ref{figseq}) for the fastest possible time resolution.
	The 9~mm inside diameter, 6.5~cm high sample chamber had a glass side wall to minimize energy transfers between grains and the side wall.
	The teflon bottom wall was pierced by an array of 30 0.46~mm diameter holes, making it possible to evacuate the sample container.
	An external vacuum gauge registered less than 200 mtorr during experiments with evacuated samples.

	A loudspeaker connected to the chamber by a fiberglass support tube and driven by a function generator and audio amplifier was used to vibrate the sample container vertically with a sinusoidal waveform.
	The amplitude and phase of the container motion were precisely measured by fitting the digitized output of a micromachined accelerometer (Analog Devices ADXL50) mounted to the support tube just below the NMR magnet.
	The gradient coils used for position (MRI) and displacement (PFG) measurements were calibrated by combining data from measurements on phantoms of known dimensions, along with diffusion measurements on a water sample.
	In this way we obtained calibration maps for each gradient coil of both the gradient in the designed direction, and the total gradient magnitude including components orthogonal to the designed direction.
	The accelerometer and vertical gradient calibrations were cross-checked by measuring the displacement waveform for a granular sample consolidated with epoxy.
	The dimensionless acceleration values $\Gamma$ reported here are accurate to within $\pm 0.1$.

\subsection{\label{samples}Granular samples}

	Previous NMR studies of granular media used plant seeds or liquid-filled capsules to take advantage of the long-lived NMR signals produced by liquids \cite{nakagawa93,knight96,caprihan97,yang00,mueth00,caprihan00}.
	For the present study, we chose mustard seeds~\cite{noteseeds} which showed no tendency to static charging.
	The seeds were ellipsoidal, with three unequal major diameters and eccentricity $\leq 20\%$.
	The average of all three diameters measured for a sample of 10 seeds was $d = 1.84$ mm with a standard deviation of 5\%, and the average mass was $m = 4.98$~mg/seed.
	These average values of $d$ and $m$ are used below for comparison with a hydrodynamic theory, which is formulated for samples of identical grains.

	To roughly measure the restitution coefficient $\epsilon$, a video camera was used to record the bounce height when the seeds were dropped onto a stone surface.
	As some linear kinetic energy is converted to rotational energy upon impact, this gives a lower bound on $\epsilon$.
	Conversely the smallest drop velocity that could readily be used (44~cm/s) was larger than the typical RMS velocity in our vibrofluidized experiments ($\leq 30$~cm/s).
	Since $\epsilon$ is often a decreasing function of impact velocity \cite{ramirez99,king02}, the restitution coefficient in the vibrofluidized experiments should be greater than is measured in the dropping trials.
	The $\epsilon$ value deduced from the highest rebound over 15 dropping trials on each of five seeds ranged from 0.83 to 0.92, with an average of 0.86.

\subsection{Data acquisition}

\begin{figure}
\includegraphics[width=1.0 \linewidth]{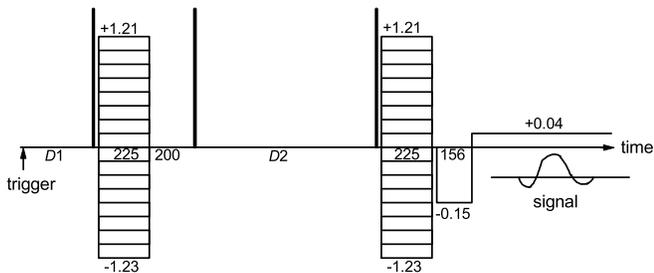}
\caption{\label{figseq}
	Pulse sequence used for PFG/MRI data acquisition.
	This is a stimulated-echo PFG sequence followed by one-dimensional frequency-encoded MRI.
	The three solid vertical bars are $\frac{\pi}{2}$ RF pulses, the boxes with horizontal hatching are gradient pulses in the displacement-resolution (PFG) direction, and the open boxes are gradient pulses in the position-resolution (MRI) direction.
	The gradient pulses are labeled with the gradient strengths in T/m, while numbers on the time axis indicate durations of intervals in microseconds.
	The start of the sequence was triggered at a fixed phase of the sample vibration, and the delay time $D1$ was varied over the range 1-20~ms to select the time in the vibration cycle at which NMR data was acquired.
	The displacement measurement time $\Delta t$ is equal to the time between PFG gradient pulses plus one third the duration of the gradient pulses; for the sequence shown $\Delta t = D2 + 380$~$\mu$s.
	An eight-sequence phase cycle was used to suppress undesired coherences and to enable acquisition of both real and imaginary parts of the complex PFG propagator.
}\end{figure}

	Proton NMR signals from the mustard-seed sample were acquired using the PFG/MRI sequence shown in Fig.~\ref{figseq}.
	After Fourier-transformation with respect to time, this sequence yields the position-dependent PFG signal (one-dimensional image)
\begin{equation}\label{hqt}
h(q_u,v) = \int p(\Delta u, v)  e^{i q_u \Delta u} dq_u 
\end{equation}
where $v=x$, $y$ or $z$ is the direction of the MRI gradient and $u=x$, $y$ or $z$ is the direction of the PFG gradient.
	In this equation $q_u$ is proportional to the area of the PFG gradient pulse, which is varied through an array of regularly spaced values (Fig.~\ref{figseq}).
	Here the ``propagator'' $p(\Delta u, v)$ is the joint probability that a grain (more precisely an oil molecule in a grain) is located at position $v$ and moves a distance $\Delta u$ during the time interval $\Delta t$ set by the pulse sequence \cite{callaghan91}.
	From Eq.~\ref{hqt}, the propagator $p(\Delta u, v)$ can be obtained by Fourier transforming the NMR signal $h(q_u,v)$ with respect to $q_u$.
	With the gradient values used in these experiments the propagator was obtained with resolutions of 500~$\mu$m in the position variable $v$ and 42~$\mu$m in the displacement variable $u$.

	This type of NMR experiment does not determine the positions or displacements of individual grains.
	Rather, an ensemble average, the propagator, is obtained.
	Provided that the propagator has a well defined time average at each phase of the vibration cycle, the NMR acquisition can be repeated as many times as necessary to build up good signal/noise.
	Allowing full longitudinal relaxation between acquisitions (the simplest but not the fastest way to take data), the acquisition time for a time-resolved propagator $p(\Delta u, v, t)$ was 21~hours.

\subsection{\label{measrms}Measurement of RMS displacement}

	For some purposes, the full propagator $p(\Delta u, v)$ is of interest, while for other purposes (e.g. measuring the granular temperature) only the position-dependent RMS displacement is required,
\begin{equation}\label{durms}
\Delta u_\text{rms}(v) = \left( \frac{1}{n(v)}\int p(\Delta u, v) (\Delta u)^2 d(\Delta u) \right) ^{1/2}.
\end{equation}
	Here $n(v) = \int p(\Delta u, v) d(\Delta u)$ is the position dependent number density.
	Both $n(v)$ and $\Delta u_\text{rms}(v)$ can be obtained directly from the NMR data (before it is Fourier transformed to yield the propagator) using
\begin{eqnarray}\label{durms2}
n(v) & = & h(0,v), \\
\Delta u_\text{rms}(v) & = & \left( -\frac{1}{n(v)} \left. \frac{\partial^2 h(q_u,v)}{\partial q_u^2} \right| _{q_u=0} \right) ^{1/2}. \nonumber
\end{eqnarray}
	For optimum signal/noise we evaluated the RHS of these expressions by fitting the central (small $|q_u|$) portion of $h(q_u,v)$ to a function including constant, quadratic, and higher terms.

\section{Experimental Results: Density and Displacement Distributions}

	In the present experiments, all measured quantities depended strongly on height $z$, due to gravity and the localization of energy injection at the sample bottom.
	Therefore, most data were taken with the MRI axis $v=z$.
	As both vertical and horizontal grain motion were of interest, for the PFG axis both $u=z$ and $u=x$ were used.
	Thus, the primary experimental data shown in this paper are the height-dependent number density $n(z)$, the vertical and horizontal displacement propagators $p(\Delta z,z)$ and $p(\Delta x,z)$, and the vertical and horizontal RMS displacements $\Delta z_\text{rms}(z)$ and $\Delta x_\text{rms}(z)$.
	In addition, some data were taken using a two-dimensional MRI sequence to verify that the grain density was approximately uniform in the horizontal plane and to calibrate the number of layers $N_\ell$ (Appendix~\ref{horizontal}).

	The granular state with $N_\ell=3.0$, $\Gamma=15$, and sample cell evacuated was selected as a reference state measured in greatest detail, and other states were selected by varying a single parameter from the reference value.
	The following variations were carried out: (a) increase of ambient air pressure up to 1~atm, (b) variation of $N_\ell$ from 2.2 to 4.0, (c) variation of $\Gamma$ up to 18.
	The upper limit on $\Gamma$ was set by the apparatus, while the lower limit on $\Gamma$ and the upper limit on $N_\ell$ were set by the requirement that the sample fully and consistently fluidize, when observed visually outside of the NMR apparatus.
	For smaller $\Gamma$ or larger $N_\ell$ we observe a complex, hysteretic transition to a partially fluidized, partially jammed state which is beyond the scope of the present study.

\subsection{\label{air}Effects of Interstitial Air}

\begin{figure}
\includegraphics[width=1.0 \linewidth]{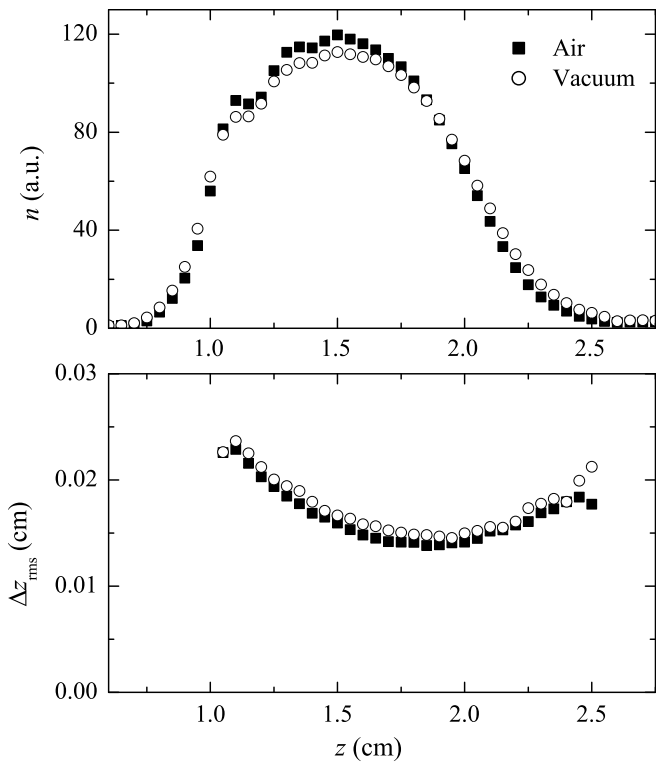}
\caption{\label{figairvac} Effect of cell evacuation.
      Un-normalized number density (top) and mean-square vertical displacement in time interval $\Delta t = 1.38$~ms (bottom) as functions of height $z$ for $N_\ell=3.0$ layers of mustard seeds vibrated at $\omega/2 \pi = 50$~Hz and acceleration amplitude relative to gravity $\Gamma=15$.
     The data labeled Air were taken in normal atmospheric conditions, while for the data labeled Vacuum the cell was evacuated so that an external pressure gauge registered less than 200~mtorr.
      The small differences between the Air and Vacuum data are comparable to the typical experimental repeatability and therefore are not significant.
}\end{figure}

	Most of our data were taken with the sample container evacuated.
	Some of the measurements were repeated in normal atmospheric conditions, to determine how large the interstitial gas effects could be.
	Experimentally, the effect of air is unobservably small (Fig.~\ref{figairvac}).
	This agrees with the following rough estimate of viscous energy losses due to air drag.
	The mean free path varies widely with height in our system but a typical value is $\lambda \approx 0.3 d$ (Appendix~\ref{mfp}).
	Similarly, a typical grain velocity is on the order of 20~cm/s.
	Using these values and the drag force on a sphere in an unbounded medium, we compute that the fraction of kinetic energy lost to viscous drag over one mean free path is $9 \times 10^{-4}$ for $p=1$~atm and $3 \times 10^{-4}$ for $p=200$~mtorr.
	These energy losses are negligible compared to the fraction of kinetic energy lost in an intergrain collision, $1-\epsilon^2 \approx 0.24$.

\subsection{Time-resolved density and displacement profiles}

\begin{figure}
\includegraphics[width=1.0 \linewidth]{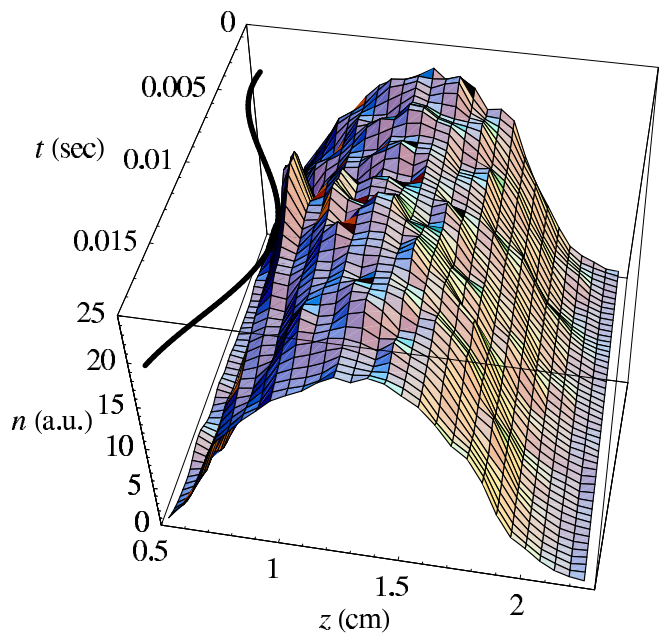}
\caption{\label{fignzt}
      Un-normalized number density of grains $n$ as a function of height $z$ and time in the vibration cycle $t$ for the reference vibrofluidized state $N_\ell=3.0$, $\Gamma=15$.
      The thick black curve shows the container vibration waveform, obtained by fitting the digitized accelerometer output.
      For this figure the curve is drawn to show the motion of the container bottom, which impacts the sample bottom at $t \approx 12$~ms.
      Apart from the disturbance caused by this impact, the density profile is largely independent of $t$.
}\end{figure}

\begin{figure*}
\includegraphics[width=1.0 \linewidth]{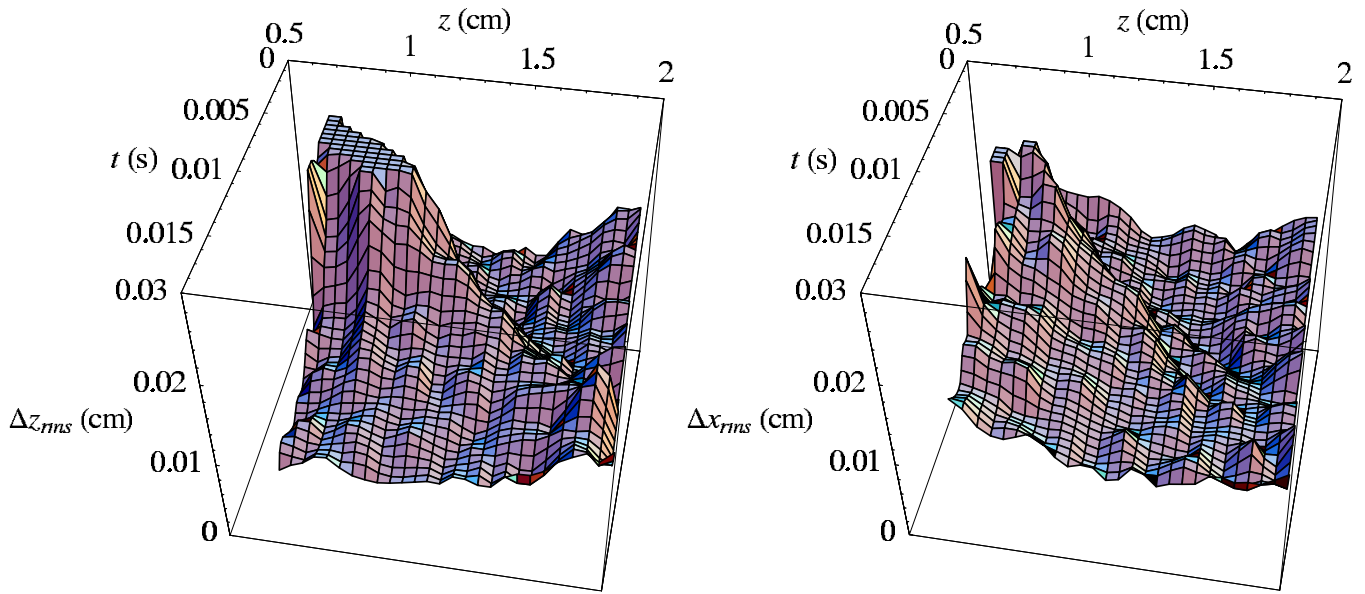}
\caption{\label{figdzxt}
	Profiles  of the vertical (left) and horizontal (right) RMS grain displacement $\Delta z_\text{rms}, \Delta z_\text{rms}$ in a time interval $\Delta t = 1.38$~ms, as functions of the height $z$ and the time in the vibration cycle $t$.
	The experimental conditions are the same as for Fig.~\ref{fignzt}.
}\end{figure*}

	In Sec.~\ref{hydrofit} below the NMR data are fitted to a hydrodynamic theory which models the granular system as a fluid with time-independent (but spatially varying) grain velocity distribution.
	In this picture, the vibrating container bottom serves to inject energy (granular ``heat'') into the system.
	For such a picture to make sense, the overall state of the granular system must be nearly independent of time within the vibration cycle.

	We have taken some NMR data as a function of $t$ within the 20~ms container vibration period, Figs.~\ref{fignzt} and \ref{figdzxt}.
	The density profile (Fig.~\ref{fignzt}) is indeed largely independent of $t$ over most of the sample, except for a noticeable disturbance at small $z$ for $t$ near the top of the container's motion, when the container floor impacts the sample.
	In the bottom layer of the sample, this disturbance damps out (thermalizes) over a time span of less than 10 ms, and above the bottom layer it is hardly evident at all.
	Similarly, in the RMS displacement profile (Fig.~\ref{figdzxt}) a large transient is visible in the vertical displacement due to grains impacted by the container bottom, while the horizontal displacement transient is much smaller relative to the time-average value.

	Over most of the cycle the sample ``floats'' well above the vibrating container floor (Fig.~\ref{fignzt}), illustrating two important points about the vibrofluidization process:
	(a)~It is possible to achieve a nearly time-independent fluid-like granular state even though the driving vibration amplitude is much greater than typical stochastic motion within the sample.
	The required conditions are (i)~$\omega^* = \omega(d/g)^{1/2} > 1$, which ensures that the sample itself only falls a grain-scale distance in one vibration period no matter how large the container oscillation amplitude is, and (ii)~$X = N_\ell(1-\epsilon) \alt 1$, allowing the sample to fully fluidize.
	For the experiments described here $\omega^* =4.3$ and $0.29 \leq X \leq 0.52$.
	(b)~For realistic multilayer granular systems, the primary contact between the grains and the vibrating cell bottom occurs near the top of the vibration stoke, when the cell bottom velocity is much less than its RMS value.
	Therefore, the energy transfer into the granular medium cannot be estimated, even roughly, from the RMS driver velocity.
	This is because the energy transfer depends upon the mean cell bottom velocity \emph{during grain impacts}~\cite{mcnamara98}.

	Apart from Figs.~\ref{fignzt} and \ref{figdzxt} the data shown in this paper have all been averaged over the vibration cycle for improved signal/noise.
	It should be born in mind that the state of the system is not, in fact time-independent very near the sample bottom and that the complex process of energy transfer from the vibrating base is not captured by these time averages.

\subsection{Short time displacements: Crossover to the ballistic regime}

\begin{figure*}
\includegraphics[width=1.0 \linewidth]{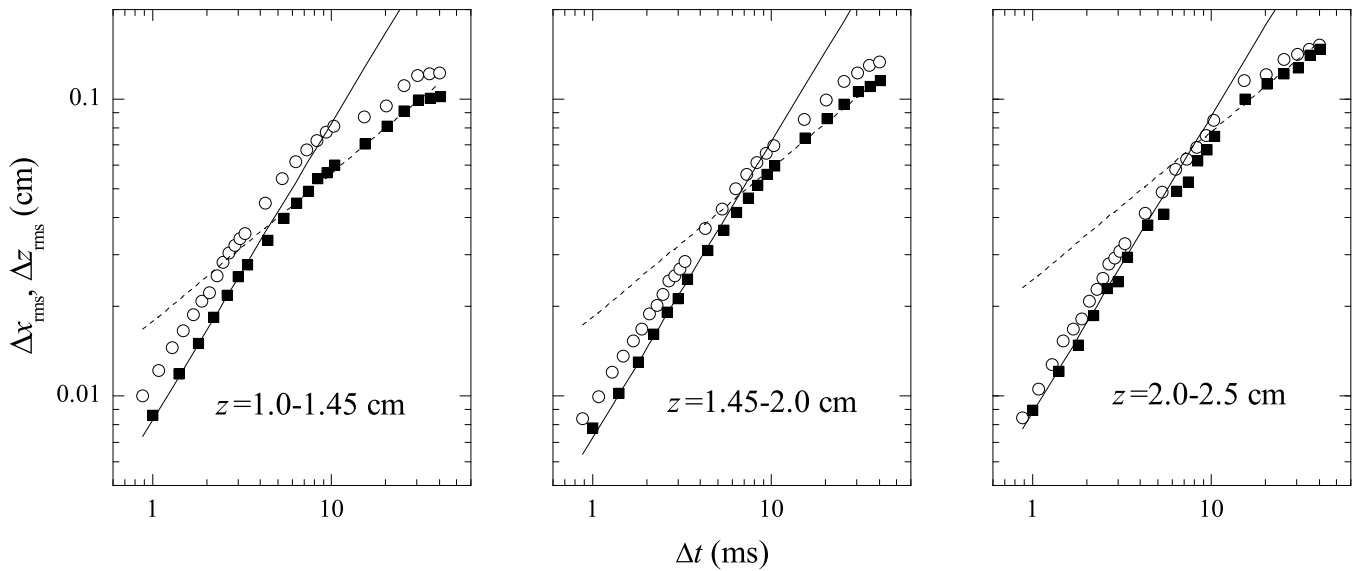}
\caption{\label{figzxrmst} Root-mean-square vertical (circles) and horizontal (squares) grain displacements $\Delta z_\text{rms}, \Delta x_\text{rms}$ versus the displacement measurement time $\Delta t$ for a sample with $N_\ell=3.0$, $\Gamma=15$.
	The data are shown for three different horizontal slices ($z$ ranges), which may be correlated with the density profile in Fig.~\ref{fignzt}.
	In each log-log plot, the solid line has a slope of unity (RMS displacement proportional to time) while the dashed line has a slope of $\frac{1}{2}$.
}\end{figure*}

	Figure~\ref{figzxrmst} shows the RMS vertical and horizontal grain motion in the reference state $N_\ell=3.0$, $\Gamma=15$ as functions of the displacement measurement time $\Delta t$, for slices at three different heights $z$ within the sample.
	At each height there is a crossover at short $\Delta t$ to the power law $\Delta z_\text{rms}, \Delta x_\text{rms} \propto \Delta t$, which indicates ballistic behavior (displacements proportional to time).
	For longer $\Delta t$ the data fall below the ballistic power law, possibly crossing over to a diffusive power law $\Delta z_\text{rms}, \Delta x_\text{rms} \propto (\Delta t)^{1/2}$.
	The crossover time, which is an estimate of a typical grain collision time, ranges from 5~ms in the dense, lower portion of the sample to 10~ms in the less dense, upper portion of the sample.
	In Appendix~\ref{mfp}, it is shown that the crossover time agrees well with the grain collision time computed from the mean free path and the measured granular temperature.

	A key feature of our experiments is the ability to take data for displacement intervals $\Delta t$ much less than the grain collision time, i.e. deep within the ballistic regime.
	The RMS displacement data shown in the remainder of this paper was taken with $\Delta t = 1.38$~ms, well within the ballistic regime for all values of $N_\ell, \Gamma$ used.

\subsection{Horizontal and vertical velocity distributions: Deviations from Gaussian}

\begin{figure*}
\includegraphics[width=0.9 \linewidth]{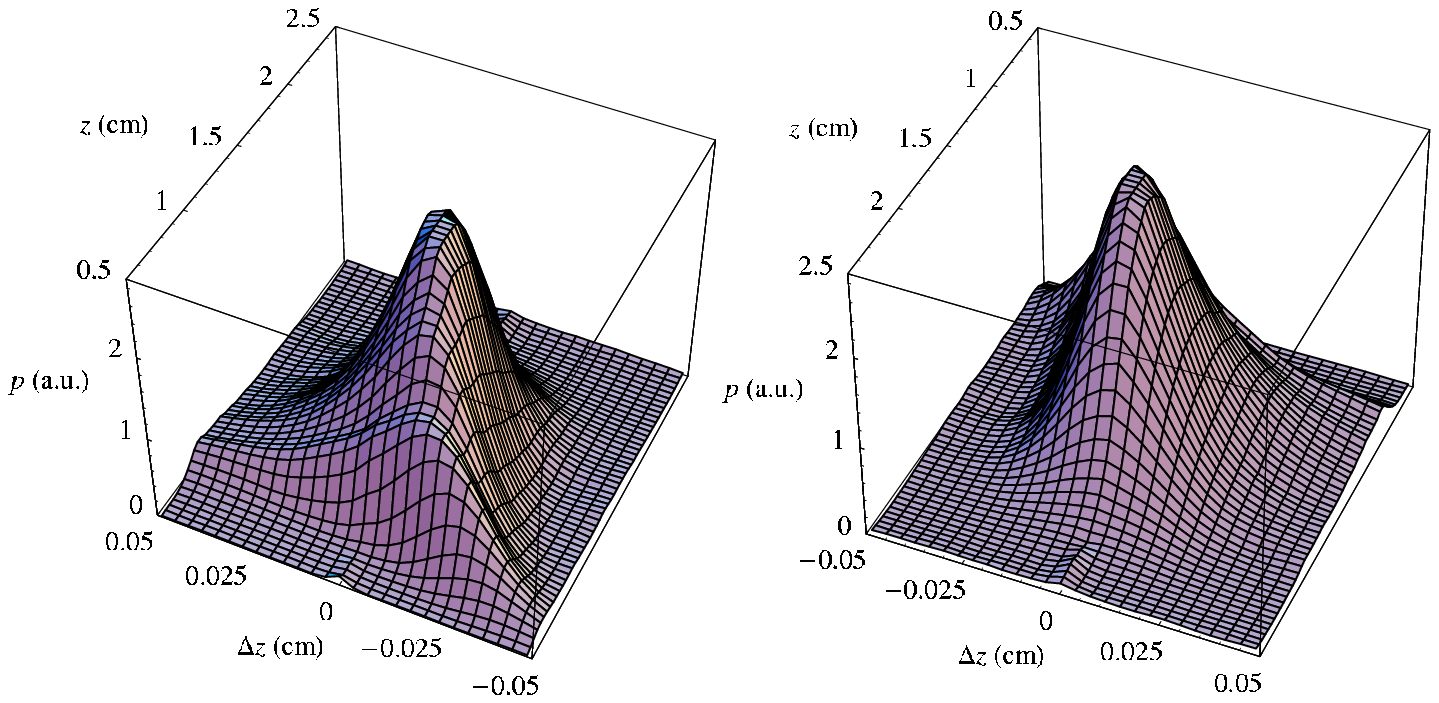}
\caption{\label{figpdzz}
	Propagator, or distribution $p(\Delta z, z)$ of vertical displacements $\Delta z$ during time interval $\Delta t = 1.38$~ms, as a function of height $z$ for the reference vibrofluidized state $N_\ell=3.0, \Gamma=15$.
	As $\Delta t$ is well within the ballistic regime, $p(\Delta z, z)$ approximates the distribution of vertical velocities $p(v_z, z)$ with $v_z = \Delta z / \Delta t$.
	The same data are shown from two different viewpoints in the left and right plots.
	Near the bottom of the sample (small $z$, most visible in the left plot) the velocity distribution is highly skewed, with a wide tail on the $v_z>0$ side due to particles traveling upward with large velocity after impacting the vibrating container bottom.
	Well away from the bottom (right plot) the velocity distribution is nearly symmetric.
}\end{figure*}

\begin{figure*}
\includegraphics[width=0.9 \linewidth]{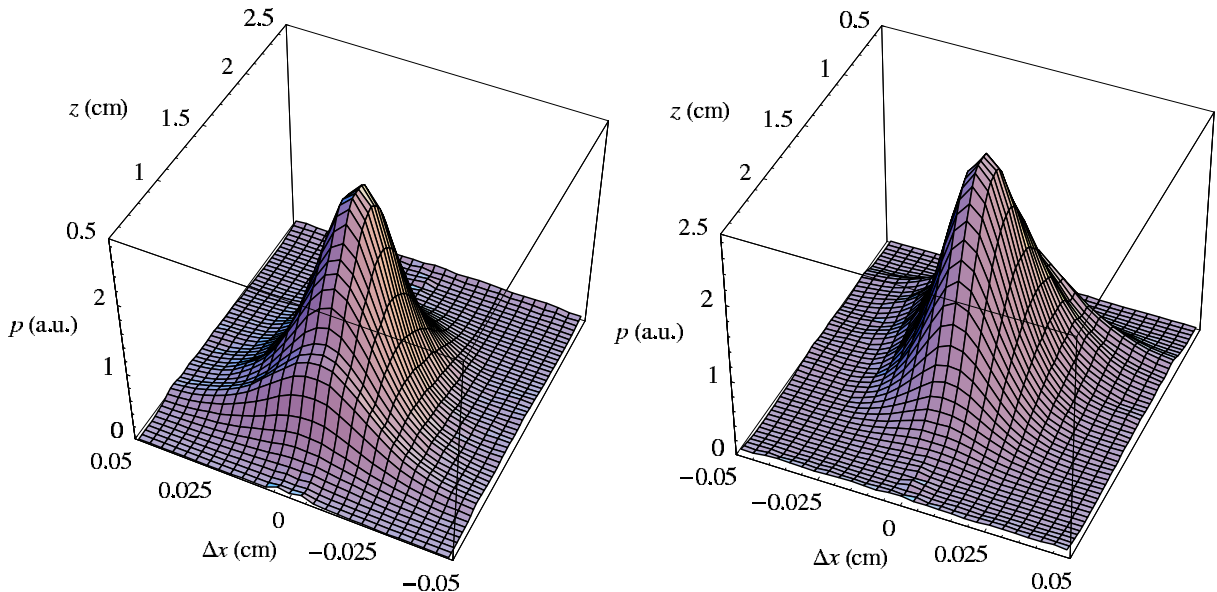}
\caption{\label{figpdxz}
	Two views of the propagator $p(\Delta x, z)$ for horizontal displacements $\Delta x$ in the same conditions as in Fig.~\ref{figpdzz}.
	The distribution of horizontal velocities is symmetric at all heights $z$, consistent with the symmetry of the experimental arrangement.
}\end{figure*}

\begin{figure}
\includegraphics[width=1.0 \linewidth] {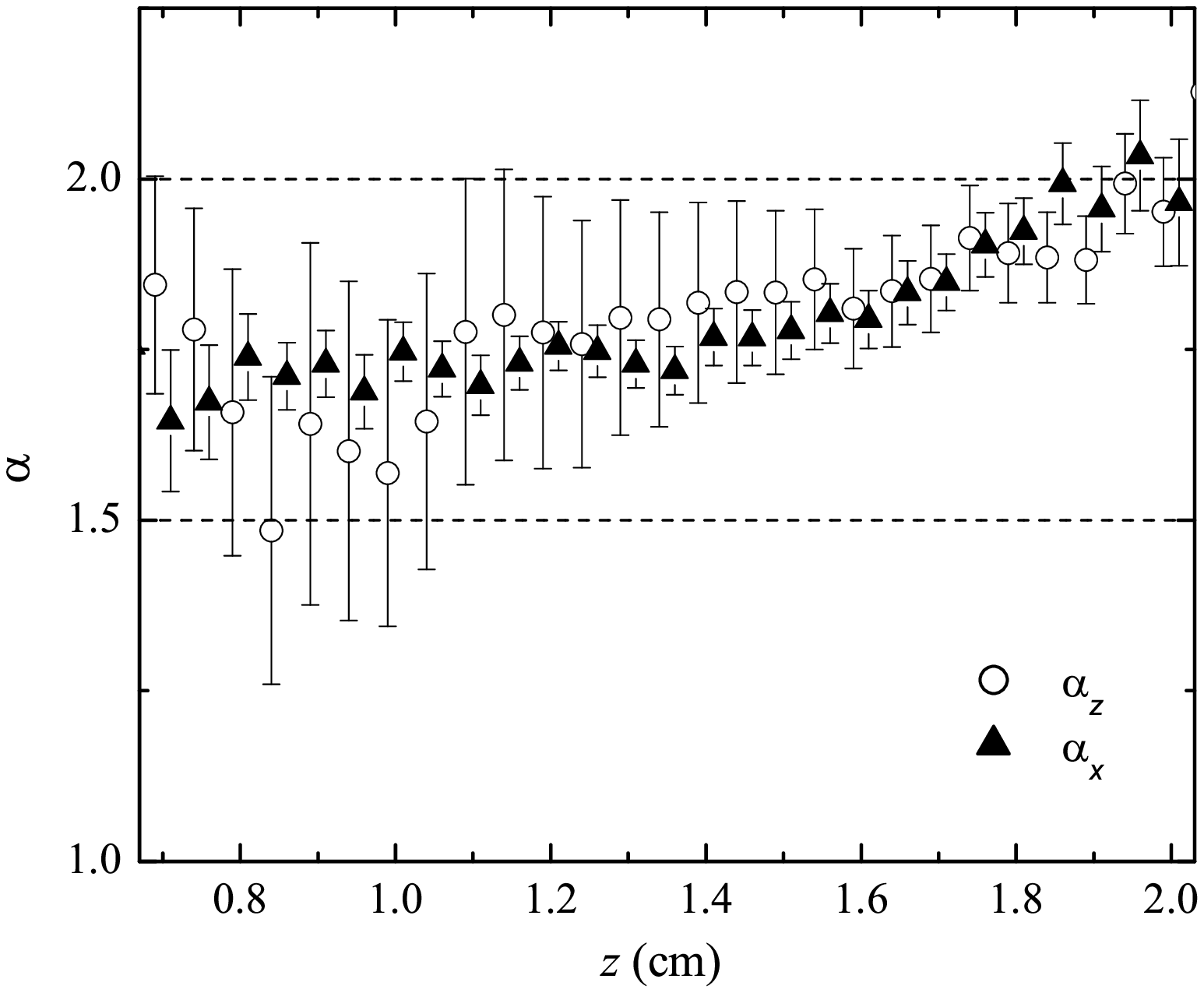}
\caption{\label{figalpha}
	Fitted stretching exponents $\alpha_{z,x}$ for the vertical and horizontal velocity distributions versus height $z$, for $N_\ell=3.0, \Gamma=15$.
	The Gaussian and $\exp(-v^{3/2})$ cases are shown by the dashed lines.
	The large uncertainties for $\alpha_z$ at small $z$ indicate a poor fit of the stretched-exponential function, which is symmetric, to the highly asymmetric velocity distribution (Fig.~\ref{figpdzz}).
}\end{figure}

	Figures \ref{figpdzz} and \ref{figpdxz} show the distributions of vertical and horizontal displacements, respectively, for the reference vibrofluidized state $N_\ell=3.0$, $\Gamma=15$ and displacement observation time $\Delta t = 1.38$~ms within the ballistic regime.
	These propagators should be good approximations to the distributions of vertical and horizontal velocities, $p(v_{z,x},z)$ with $v_{z,x}=\Delta z,x/\Delta t$.

	In some previous experiments \cite{losert99,olafsen99,rouyer00,blair01} and simulations \cite{brey02} of granular fluids (mostly in reduced dimensions) significant deviations have been found between the measured velocity distribution, and the Gaussian (Maxwell-Boltzmann) distribution that describes an elastic fluid in equilibrium, $p(v_i) \propto \exp(-mv_i^2/2T)$ (setting Boltzmann's constant to unity).
	For example, Rouyer and Menon \cite{rouyer00} found that the distribution of horizontal velocities in a two-dimensional vibrofluidized media robustly obeyed
\begin{equation}\label{alpha}
	p(v_x) \propto \exp[-|v_x/\sigma_\alpha|^\alpha]
\end{equation}
with $\alpha=1.55 \pm 0.1$.
	Using kinetic theory it has been shown that the \emph{tails} of the velocity distribution should follow Eq.~\ref{alpha} with $\alpha=1$ for a freely-cooling granular gas, and $\alpha=3/2$ for a granular gas with stochastic forcing applied to every grain \cite{vannoije98a}.
	However, for the freely-cooling case the \emph{central} part of the distribution is nearly Gaussian ($\alpha=2$) and the crossover to $\alpha=1$ occurs very far out in the tails \cite{brey99}.

	In the present experiments and those of Ref.~\onlinecite{rouyer00} a steady state was achieved by supplying energy from the boundaries of the system, which corresponds neither to the freely cooling nor to the uniformly heated case.
	A few recent theoretical and simulation studies have treated the two-dimensional granular fluid with a strong thermal gradient, as occurs in vibrofluidization experiments, and have found features such as anisotropic granular temperature, asymmetric velocity distributions, and non-universal values of the velocity-stretching exponent $\alpha$ \cite{sunthar00,brey01a,brey02,barrat02}.

	In previous work the measured velocity distribution typically has been scaled by the RMS velocity $\sigma$ (note $\sigma \neq \sigma_\alpha$).
	This is possible for simulations and for video experiments in which every particle is tracked, allowing $\sigma$ to be computed unambiguously.
	In NMR experiments there is a noise floor, visible in Figs.~\ref{figpdzz},\ref{figpdxz}, which could conceal tails of the distribution contributing to $\sigma$.
	One possibility would be to use direct measurement of the RMS velocity from $q$-space data, described in Sect.~\ref{measrms} and used below to measure the granular temperature.
	Here we have taken the simpler approach of directly fitting the experimental velocity distributions to Eq.~\ref{alpha}, allowing the width parameter $\sigma_\alpha$ to vary.
	The results of these fits are shown in Fig.~\ref{figalpha}.

	The fitted stretching exponents for vertical and horizontal velocity distributions $\alpha_z, \alpha_x$ generally lie between 1.5 and 2.0, with values closer to 2.0 (Gaussian) near the top of the sample.
	Altough $\alpha_z$ and $\alpha_x$ are not significantly different, the large error bars for $\alpha_z$ near the sample bottom reflect the fact that the experimental velocity distribution is asymmetric (Fig.~\ref{figpdzz}) and hence does not fit Eq.~\ref{alpha} well.
	The present results for $\alpha_x$ in a three-dimensional vibrofluidized system appear marginally different from the corresponding two-dimensional results of Ref.~\onlinecite{rouyer00}.
	Some of this difference may simply be due to the different way in which we fit the data, allowing $\sigma_\alpha$ to vary rather than fixing it to a precisely measured experimental value.
	Our results are similar to the two-dimensional simulation results of Ref.~\cite{brey02}, which found deviations from Gaussian are most pronounced near the bottom of the system.

\section{\label{hydrofit}Fit to a Hydrodynamic Theory}

\subsection{Granular hydrodynamics for a stationary vibrofluidized state}

	Granular hydrodynamic equations have been derived by many groups, using increasingly sophisticated approximation schemes \cite{jenkins83,haff83,lun84,grossman97,vannoije98,sela98,garzo99}.
	In this section we show how any of these theories may be numerically integrated for comparison with experiments.
	Section \ref{gzfit} shows this comparison between the theory of Ref.~\onlinecite{garzo99} and our experimental results.

	In Ref.~\cite{bougie02} granular hydrodynamics was successfully used to describe a \emph{nonstationary} simulation with large variations of the density profile over the course of a vibration cycle.
	Here, we specialize to the simpler situation in which the average velocity field $\mathbf{v}(\mathbf{r},t)$ vanishes and the number density field $n(\mathbf{r},t)$ and temperature field $T(\mathbf{r},t)$ are independent of $t$ and depend upon $\mathbf{r}$ only via the height coordinate $z$.
	We consider a system of $N$ identical spherical grains of diameter $d$ and mass $m$.

	With these assumptions the pressure $p(z)$ and number density $n(z)$ satisfy a hydrostatic equilibrium equation
\begin{equation}\label{dpdz}
	dp/dz = -mgn
\end{equation}
and the upward heat flux $q_z(z)$ obeys an energy flux balance equation
\begin{equation}\label{dqdz}
	dq_z/dz = -P_c
\end{equation}
	where $P_c$ is rate of kinetic energy loss per unit volume, due to inelastic collisions.
	The heat flux is
\begin{equation}\label{qz}
	q_z = -\kappa (dT/dz) -\mu (dn/dz)
\end{equation}
where $\kappa$ is the conventional thermal conductivity, and $\mu$ is a transport coefficient that is nonzero only for inelastic systems \cite{lun84}.
	Due to the second term in Eq.~\ref{qz}, it is possible to have a flow of heat from colder regions to warmer ones.
	Although a number of authors have calculated $\mu$ \cite{lun84,sela98,garzo99} and discussed the consequences of $\mu \neq 0$ \cite{soto99,brey01,ramirez02}, to our knowledge no heuristic explanation has been proposed for the existence of a heat current driven by density gradients.
	This density-driven heat current has a marked effect in the present experiments.

	Equations \ref{dpdz}-\ref{qz} are simply the results of definitions and of momentum and energy conservation.
	A hydrodynamic theory closes this set of equations by providing expressions for $p(n,T)$ (the equation of state), $P_c$, $\kappa$ and $\mu$.
	It is convenient to nondimensionalize variables using the grain diameter $d$, grain mass $m$, and gravity $g$,
\begin{eqnarray}\label{dimless}
	z^* &=& (z-z_0)/d, \\
	n^* &=& nd^3, \nonumber\\
	T^*_{(i)} &=& T_{(i)}/mgd, \nonumber\\
	q_z^* &=& q_z (d/g)^{3/2}/m. \nonumber
\end{eqnarray}
	Here $z_0$ is the height of the bottom of the sample, corresponding to $z^*=0$.

	For a hard-sphere system with constant restitution coefficient the equation of state takes the form
\begin{equation}\label{eos}
	p/nT = f(n^*).
\end{equation}
	Using Eqs.~\ref{dpdz}-\ref{qz} and \ref{eos} the following expression is derived for the density gradient:
\begin{equation}\label{dndz}
	\frac{dn}{dz} = \frac{q_z/\kappa - mg/f}
{(T/n)(1 + \frac{n^*}{f} \frac{df}{dn^*}) - \mu/\kappa}.
\end{equation}

	Equations~\ref{dpdz}, \ref{dqdz} and \ref{dndz} are coupled first order ordinary differential equations for $\{p,q_z,n\}$ that can be numerically integrated from the sample bottom at $z_0$ to $z=+\infty$.
	The boundary conditions are as follows.
	For a system in gravity with no upper wall, the pressure and heat current satisfy $p(+\infty)=0$, $q_z(+\infty)=0$.
	Using Eq.~\ref{dpdz} this gives $p(z_0)=mgn_A$, where $n_A = \int_{z_0}^{+\infty}n(z)dz$ is the total number of grains in the sample per unit horizontal cross section.
	The heat current into the system at the bottom $q_z(z_0)$ is an input parameter determined by the vibration amplitude and frequency and the system response.
	For given values of $q_z(z_0)$ and $p(z_0)$, the density at the bottom $n(z_0)$ must be adjusted to satisfy the upper boundary condition $q_z(+\infty)=0$.

	Using our convention, the dimensionless (starred) forms of Eqs.~\ref{dpdz}-\ref{dndz} and the boundary conditions are obtained by setting $m=g=d=1$.
	We have carried out the numerical integrations and present comparisons with the experiments in this dimensionless form.

\subsection{\label{gzfit}Fit of the experimental results to the Garz\'{o}-Dufty hydrodynamic theory}

\begin{figure}
\includegraphics[width=1.0 \linewidth]{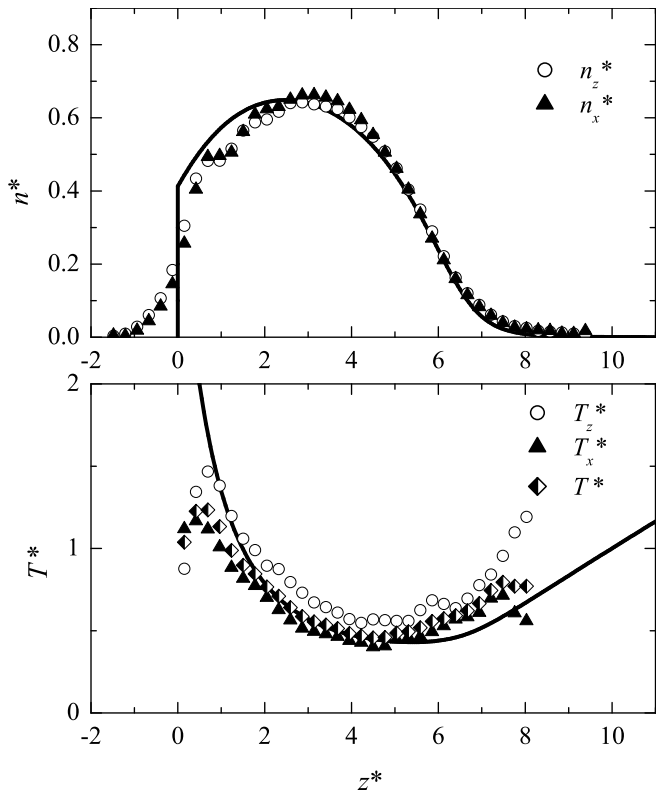}
\caption{\label{fign60g15}
	Experimental density and granular temperature profiles (symbols) and fit to the hydrodynamic theory of Ref.~\onlinecite{garzo99} (curves) for $N_\ell=3.0$ layers of grains vibrated at dimensionless acceleration amplitude $\Gamma=15$.
	Dimensionless variables defined in Eq.~\ref{dimless} are used.
	In the top graph the symbols labeled $n_z^*$ show the density profile $n^*(z^*)$ derived from the data set used to measure vertical displacements while the symbols labeled $n_x^*$ show the same quantity derived from the horizontal-displacement data set; these two profiles should be identical and differences reflect experimental accuracy.
	In the bottom graph $T_z^*$, $T_x^*$ are the measured granular temperature profiles for vertical and horizontal displacements, respectively, measured using observation interval $\Delta t = 1.38$~ms.
	It is expected that $T_z^* > T_x^*$, at least near the bottom of the sample, as energy is input only to the vertical degrees of freedom by the contain vibration.
	The theory should be compared with the symbols labeled $T^*$, which are computed from the measured temperatures using $T^* = \frac{1}{3}T_z^* + \frac{2}{3}T_x^*$.
}\end{figure}

\begin{figure}
\includegraphics[width=1.0 \linewidth]{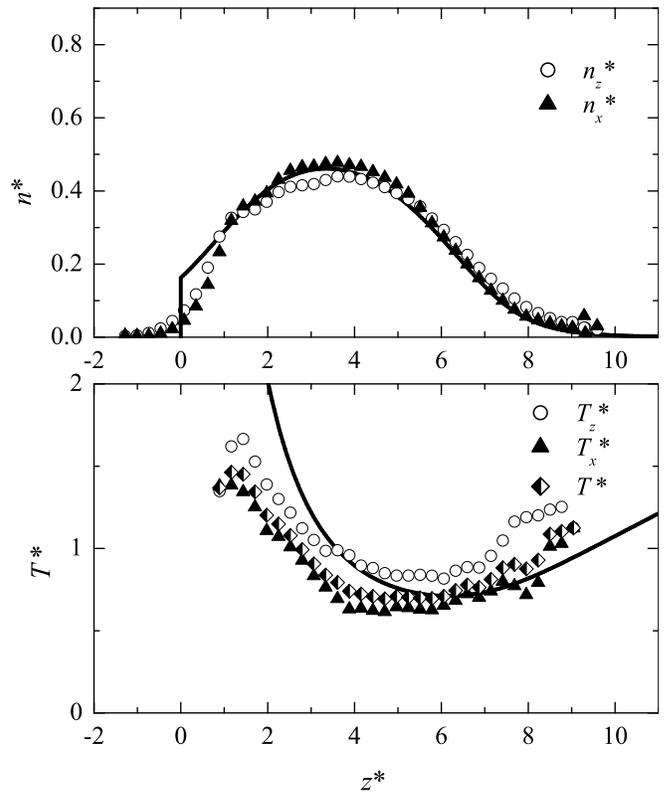}
\caption{\label{fign45g15}
	Density and temperature profiles for $N_\ell=2.2$, with other conditions the same as in Fig.~\ref{fign60g15}.
	Under these conditions, a significant discrepancy can be seen between the experimental and theoretical granular temperatures in the lower portion of the sample, $z^* \alt 3$.
	We observe this discrepancy in conditions such that the density in the lower portion of the sample is well below the peak density (upper graph).
}\end{figure}

\begin{figure}
\includegraphics[width=1.0 \linewidth]{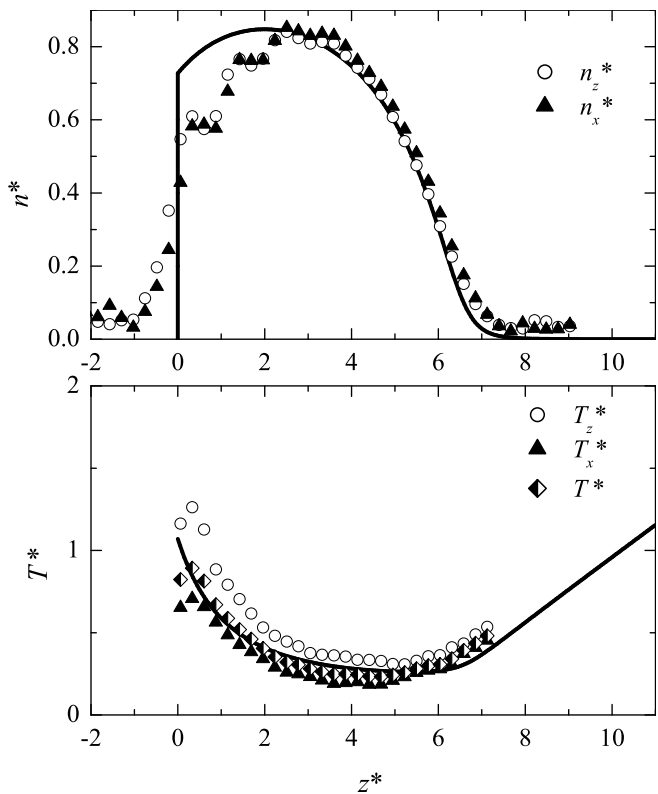}
\caption{\label{fign80g15}
	Density and temperature profiles for $N_\ell=4.0$, with other conditions the same as in Fig.~\ref{fign60g15}.
	The peak density $n^*=0.85$ observed in these conditions corresponds to volume-filling fraction $\phi = (\pi/6)n^* = 0.45$, about 75\% of the random close packed value.
}\end{figure}

\begin{figure}
\includegraphics[width=1.0 \linewidth]{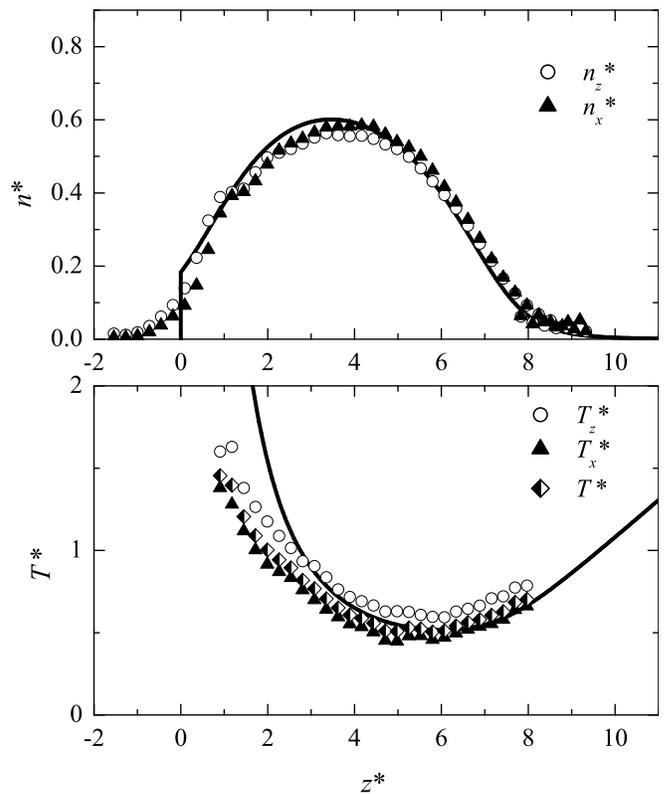}
\caption{\label{fign60g18}
	Density and temperature profiles for $\Gamma=18$, with other conditions the same as in Fig.~\ref{fign60g15}.
	As in Fig.~\ref{fign45g15} the density at the bottom of the sample is much less than the peak density, and the measured granular temperature does not agree well with the theoretical granular temperature near the sample bottom.
	Note that this situation is achieved here by increasing $\Gamma$ relative to the reference state, while in Fig.~\ref{fign45g15} it was achieved by decreasing $N_\ell$.
}\end{figure}

	In this section, we show fits of our experimental data to a formulation of granular hydrodynamics by Garz\'{o} and Dufty \cite{garzo99}.
	This calculation is based on the ``revised Enskog theory'', which is accurate for \emph{elastic} fluids over a wide range of densities and is extended in Ref.~\onlinecite{garzo99} to arbitrary values of the restitution coefficient $\epsilon$.
	Some other theories of granular hydrodynamics are derived as expansions in powers of $1-\epsilon^2$ \cite{sela98}, and thus may only be accurate for $\epsilon \approx 1$.
	On the other hand, the Garz\'{o}-Dufty theory makes the molecular chaos assumption that the pre-collision grain velocities are uncorrelated.
	Pre-collision velocity correlations are known to exist in inelastic hard-sphere fluids \cite{soto01,pagonabarraga01}, and can be treated theoretically using ``ring kinetic theory'' \cite{vannoije98}.

	It is doubtful whether fits to the present data could distinguish between these various hydrodynamic theories.
	We chose to compare our data to Ref.~\onlinecite{garzo99} because this calculation is nominally valid for wide ranges of density and inelasticity, and because the theoretical results for $f$, $P_c$, $\kappa$ and $\mu$ are presented in an algebraic form convenient for numerical integration.
	In this theory the granular temperature is assumed to be isotropic, $T_x = T_y = T_z$, where $T_i = m \langle v_i^2 \rangle$, but it is not assumed that the velocity distribution is Gaussian.
	Experimentally we find modest differences (of order 20\% or less) between the horizontal and vertical granular temperatures, $T_{x,y} \neq T_z$, and more generalized formulations of hydrodynamics may be able to treat this anisotropy \cite{sunthar00,brey01a,dufty01}.
	For comparison with the isotropic-temperature theory the measured vertical temperature $T_z$ and horizontal temperature $T_x$ are averaged using $T = \frac{1}{3}T_z + \frac{2}{3}T_x$.

	Figures~\ref{fign60g15} through \ref{fign60g18} show a joint fit to the Garz\'{o}-Dufty theory of our data for three different values of the dimensionless bed height $N_\ell$ and two different values of $\Gamma$.
	For this fit the height of the sample bottom $z_0$ was allowed to be different for each data set, since the sample ``floats'' above the vibration center by an amount that depends upon the dynamic interplay between energy input and gravity (Fig.~\ref{fignzt}).
	Similarly, the energy input $q_z^*(0)$ was a separate fitting parameter for each data set.
	Conversely a single value of the restitution coefficient $\epsilon$ was used to fit all four data sets.
	Thus the only fitting parameters are $z_0$, $q_z^*(0)$ and $\epsilon$.
	All other quantities entering the fits shown in Figs.~\ref{fign60g15}-\ref{fign60g18} are measured experimentally, including all of the axis variables in the figures.

	The fitted value of the restitution coefficient is $\epsilon=0.87$, in reasonable agreement with the approximate measurement described in Sec.~\ref{samples}.
	The fitted values of the dimensionless heat flux at the sample bottom $q_z^*(0)$ are listed in Table~\ref{qtable}.
	The total power input is computed as $P=(\pi D^2/4)q_z$, where $D=0.87$~cm is the effective sample diameter (Appendix~\ref{horizontal}).
	By equating the rate of momentum transfer to the total sample weight $Nmg$ the input power is rigorously related to the average over grain impacts of the container-bottom velocity $\langle v_i \rangle$ by $P = 2Nmg\langle v_i \rangle$ \cite{mcnamara98}.
	Table~\ref{qtable} lists $\langle v_i \rangle$ derived in this way from the fitted heat current, along with the ratio of $\langle v_i \rangle$ to the peak container velocity $v_0 = \omega z_0$.

\begin{table}\caption{\label{qtable}
	Fitted dimensionless heat flux into the bottom of the sample $q_z^*(0)$ as a function of dimensionless vibration acceleration $\Gamma$ and bed depth $N_\ell$.
	Also listed are the average container impact velocity $\langle v_i \rangle$ derived from $q_z^*(0)$ and the ratio of $\langle v_i \rangle$ to the vibration velocity amplitude $v_0$.
}\begin{ruledtabular}
\begin{tabular}{ccccc}
$\Gamma$ & $N_\ell$ & $q_z^*(0)$ & $\langle v_i \rangle$ & $\langle v_i \rangle/v_0$ \\
\hline
15 & 2.2 & 3.3 & 8.7~cm/s & 0.185 \\
15 & 3.0 & 3.4 & 6.7~cm/s & 0.143 \\
15 & 4.0 & 3.5 & 5.3~cm/s & 0.110 \\
18 & 3.0 & 4.8 & 9.4~cm/s & 0.168 \\

\end{tabular}
\end{ruledtabular}
\end{table}

	The fitted power input $q_z^*(0)$ is an increasing function of $\Gamma$, as expected on physical grounds.
	The power input also increases with bed depth $N_\ell$, which is expected from the momentum-balance argument, but the increase is very slow.
	One caveat is that these $q_z^*$ values are derived from the hydrodynamic fits, which tend to fail where the density at the sample bottom is much less than the peak density (Figs.~\ref{fign45g15},\ref{fign60g18}).
	In these cases particularly, the fitted $q_x^*$ represents an \emph{effective} power input reflecting the bottom boundary condition, which is greater than the \emph{actual} power input because the hydrodynamic fit has larger granular temperature than the experimental data and hence greater collision losses in the lower part of the sample.

	One robust conclusion from Table~\ref{qtable} is that the average container-grain impact velocity is much less than the peak container velocity, $\langle v_i \rangle \ll v_0$ (substituting the actual input power for the effective power only strengthens this result).
	This conclusion corresponds to the experimental observation that the sample ``floats'' above the vibrating container bottom (Fig.~ \ref{fignzt}), so that most container-grain impacts occur near the top of the vibration stoke where the container velocity is much smaller than its peak value.

	In both the data and the hydrodynamic fits in Figs.~\ref{fign60g15}-\ref{fign60g18} there is a ``temperature inversion''---the temperature increases with height near the free upper surface of the sample.
	In the hydrodynamic theory the heat flux is strictly upwards so the temperature inversion is only made possible by the term in $\mu$ in Eq.~\ref{qz}.
	The temperature inversion may be visible in some of the earliest experiments on quasi-two dimensional systems \cite{clement91,warr95} and has been extensively discussed theoretically \cite{soto99,brey01,ramirez02}.
	The results presented here are perhaps the first demonstration that the temperature inversion in a physical system can be quantitatively explained by granular hydrodynamics.

\section{Conclusions}

	One conclusion of this work is that it is possible, under certain conditions, to quantitatively model a physical vibrofluidized granular system using granular hydrodynamics.
	This is true despite the small sizes of the systems measured in grain diameters: the systems were small horizontally due to apparatus constraints, and in any case were necessarily small vertically in order to meet the fluidization condition $X \alt 1$.
	The overall success of the hydrodynamic description, demonstrated by the fits in Figs.~\ref{fign60g15}-\ref{fign60g18}, can probably be attributed to the fact that hydrodynamics only breaks down when the mean free path becomes much larger than other relevant length scales.

	A limitation of present hydrodynamic theory is evident in these figures---the inability to accurately describe the complex process of energy transfer from the vibrating container bottom to the sample.
	The actual velocity distribution is highly skewed and anisotropic near the sample bottom (Fig.~\ref{figpdzz}), and a discrepancy develops between the theoretical and measured granular temperatures when the vibrational power input is large enough to deplete the particle density near the sample bottom (Figs.\ref{fign45g15}, \ref{fign60g18}).

	To calculate the power input, typically it has been assumed that the mean free path is much larger than the vibration stroke, or that the vibration waveform is sawtooth or triangular \cite{mcnamara98,brey01}.
	Such waveforms are unrealistic since they require infinite acceleration at the peak of the vibration stroke, which is precisely the point at which energy transfer to the grains occurs (Fig.~\ref{fignzt}).
	Similarly the mean free path is much \emph{less} than the vibration stroke in the current experiments (Appendix~\ref{mfp}), which will be true in general for dense uniformly fluidized systems unless very high vibration frequencies are used.
	One possibility for deriving the power input might be to use \emph{time-dependent} hydrodynamics to model the formation and dissipation of shocks, as in Ref.~\onlinecite{bougie02}.

	Thus it appears that current granular hydrodynamic theory can accurately describe the uniformly vibrofluidized state well away from the vibrating energy source, but that derivation of the time-averaged boundary condition for realistic vibration waveforms remains an unsolved problem.
	If that problem can be solved, a complete first-principles description of the vibrofluidized state applicable to real granular systems will be in hand.

\begin{acknowledgments}
	This work was supported by NSF Grant No. CTS 9980194.
	The authors thank N. Menon for useful discussions.
\end{acknowledgments}

\appendix

\section{\label{horizontal}Horizontal density profiles and calibration of bed height}

\begin{figure}
\includegraphics[width=1.0 \linewidth]{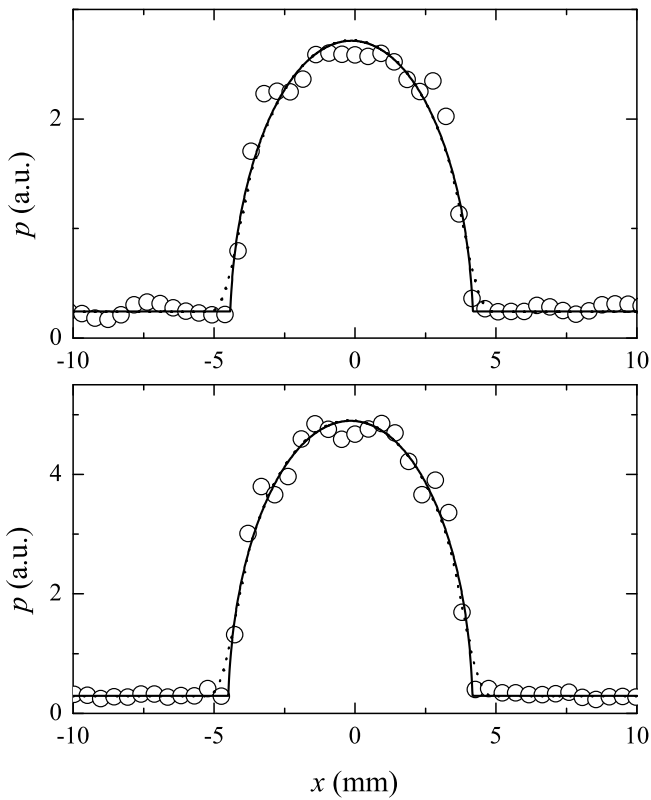}
\caption{\label{fignx}
	Horizontal density profiles $p(x,z)$ for $z=1.75$~cm (top) and $z=1.55$~cm (bottom), for the $N_\ell=3.0$, $\Gamma=15$ reference state.
	The solid curves show fits to the profile for a uniform distribution of point-source grains in a cylindrical volume.
	The dashed curves show fits assuming the grain centers are uniformly distributed, but that the NMR signal source is a sphere with the full grain diameter.
	The difference in the fitted cylinder diameter between the two fits is not significant.
}\end{figure}

\begin{figure}
\includegraphics[width=1.0 \linewidth]{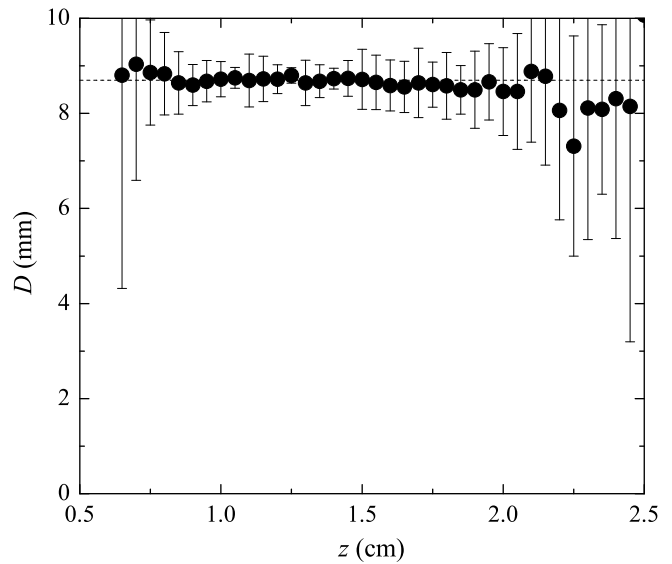}
\caption{\label{figwz}
	Fitted effective tube diameter $D$ as a function of height $z$ for the $N_\ell=3.0$, $\Gamma=15$ reference state.
	The error bars become large at the top and bottom of the sample where the density is low.
	The dashed line shows the average value $D=8.7$~mm used to analyze the data.
}\end{figure}

	The hydrodynamic model to which we have fit our data assumes that all quantities depend only upon the height coordinate $z$.
	This is not strictly true due to boundary effects at the vertical wall, as demonstrated by previous experiments \cite{wildman01b}.
	The boundary condition also affects the effective cross sectional area of the sample tube, which is needed to compute the number of grain layers $N_\ell$ for a given number of grains $N$.
	To address these issues we have taken some two-dimensional MRI data (Fig.~\ref{fignx}) to examine the horizontal density profile of the vibrofluidized system,
\begin{equation}
p(x,z) =  \int n(x,y,z) dy.
\end{equation}
	If the NMR signal from each grain comes precisely from the grain center and the grain centers are uniformly distributed horizontally over a cylinder of diameter $D$, the profile has the form
\begin{equation}\label{p0}
p(x,z) =  \left\{ \begin{array}{ll}
	f(z)[1-(2x/D)^2]^{1/2} & \text{if $x<D/2$,} \\
	0 & \text{otherwise.}
\end{array} \right.
\end{equation}
	We also considered the possibility that the NMR signal from a grain comes from a sphere of diameter $\delta \leq d$, in which case the profile of Eq.~\ref{p0} must be convoluted with the profile for a single grain.
	We find that the fitted sample diameter is not significantly changed by varying $\delta$ over the entire possible range, Fig.~\ref{fignx}.
	In this figure some systematic deviation between the data and fit is visible, hinting that the density may be slightly depressed in the sample center \cite{wildman01b}.
	However, these data are too noisy to be directly inverted to yield the radial density profile.
	The fitted diameter is consistent with a constant value $D=8.7$~mm (Fig.~\ref{figwz}).
	This is less than the actual inside diameter (9~mm) but by less than the grain diameter (1.84~mm), suggesting a slight density enhancement at the walls due to the dynamic boundary condition.

	For the fits to the hydrodynamic theory, the height of the granular bed is specified by the dimensionless number density $n_A^* = n_A d^2 = Nd^2/(\pi D^2/4)$. 
	A second measure of the bed depth is the number of layers $N_\ell$, defined as the height of the bed at rest divided by the grain diameter.
	The value of $N_\ell$ depends on the assumed volume filling fraction of the bed at rest $\phi_0$ via $N_\ell=(\pi/6\phi_0)n_A^*$. 
	For the $N_\ell$ values quoted in this paper, shown in Table~\ref{ntable}, we have used $\phi_0=0.60$ corresponding to a ``loose random packing'' \cite{dullien92}.
	It should be emphasized that the hydrodynamic fits are independent of the assumed value for $\phi_0$, as they are based on $n_A^*$ and not $N_\ell$.

\begin{table}\caption{\label{ntable}
	Three measures of the sizes of the granular samples used in this study.
The number of grains $N$ was counted precisely.
	The dimensionless area density $n_A^*$ is computed from $N$, the measured average grain diameter, and the fitted effective sample tube diameter from MRI experiments on the reference vibrofluidized state.
	The number of layers $N_\ell$ depends additionally upon the assumed volume filling fraction of the bed at rest.
	The quantity $X$ which characterizes the possibility of fully fluidizing the samples is also listed, using the value $\epsilon=0.87$ resulting from the hydrodynamic fits.
}\begin{ruledtabular}

\begin{tabular}{cccc}
$N$ & $n_A^*$ & $N_\ell$ & $X=N_\ell(1-\epsilon)$\\
\hline
45 & 2.56 & 2.2 & 0.29 \\

60 & 3.42 & 3.0 & 0.39 \\
80 & 4.56 & 4.0 & 0.52 \\
\end{tabular}
\end{ruledtabular}
\end{table}

\section{\label{mfp}The mean free path}

\begin{figure}
\includegraphics[width=1.0 \linewidth]{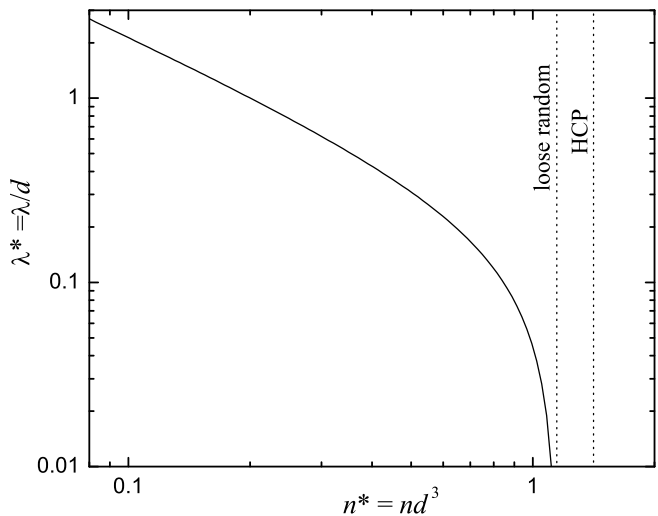}
\caption{\label{figmfp}
	Mean free path $\lambda$ as a function of number density $n$ computed from Eq.~\ref{lamstar}.
	Both quantities are nondimensionalized by the grain diameter $d$.
}\end{figure}

	For hydrodynamics to be valid, the mean free path $\lambda$ must not be too large relative to distances over which the hydrodynamic fields vary.
	Therefore it is important to estimate $\lambda$ for the experimental conditions, and this also provides a consistency check for the measured ballistic/diffusive crossover time.
	Extending the estimate of Ref.~\onlinecite{grossman97} to three dimensions gives
\begin{equation}\label{lamstar}
	\lambda^* = \lambda/d = \frac{A(n_0^*-n^*)}{n^*(B-n^*)}
\end{equation}
where $A$ and $B$ are chosen to give the exact dilute limit $\lambda^* = 1/(\pi\sqrt{2}n^*)$ for $n^* \rightarrow 0$ and the heuristic dense limit $\lambda^* = (1-n^*/n_0^*)/3$ for $n^* \rightarrow n_0^*$.
	Here $n_0^*=(6/\pi)\phi_0$ is the density at which the grains come into contact, which we arbitrarily assume occurs at the volume filling fraction for loose random packing $\phi_0 = 0.60$.

	The mean free path $\lambda^*(n^*)$ computed from Eq.~\ref{lamstar} is shown in Fig.~\ref{figmfp}.
	At the peak density $n^*=0.85$ that occurs in our experiments (Fig.~\ref{fign80g15}), Eq.~\ref{lamstar} gives $\lambda^* \approx 0.1$.
	Thus it is hardly surprising that the hydrodynamic theory can explain the spatial dependence of the density and granular temperature in the high-density regions of the sample.
	Conversely, the data fit the hydrodynamic theory well for densities extending below $n^*=0.1$, which corresponds to $\lambda^* > 2$.
	Thus, the theory continues to work for mean free paths comparable to or even somewhat greater than the scale over which the temperature and density change.

	The mean free path calculation can also be used to check the consistency of observed the ballistic crossover time, Fig.~\ref{figzxrmst}.
	For example, the center panel in this figure corresponds to an average dimensionless height $z^*=5.2$, and at this height the measured dimensionless density and temperature are $n^*=0.41$, $T^*=0.5$ (Fig.~\ref{fign60g15}).
	Using $\lambda = d \lambda^*(n^*)$ and $v_\text{rms}=(3T^*gd)^{1/2}$ we calculate~\cite{reifch12} the ballistic to diffusive crossover time $t_\text{cr} = (32/3\pi)^{1/2}\lambda/v_\text{rms} = 6.3$~ms, in excellent agreement with the observed value (Fig.~\ref{figzxrmst}).

\bibliography{vibro,granflow}

\begin{thebibliography}{61}
\bibitem[{not({\natexlab{a}})}]{noteepsilon}
\bibinfo{note}{The restitution coefficient $\epsilon < 1$ is the ratio of final
  to initial speeds for two colliding grains in their center of mass frame. In
  physical systems $\epsilon$ depends upon speed and other impact parameters,
  but many simulations and theories assume $\epsilon =$~const.}
\end{thebibliography}
\end{document}